\documentclass{article}%
\usepackage{amsmath}
\usepackage{amsfonts}
\usepackage{amssymb}
\usepackage{graphicx}%
\begin{document}

\title{An intermediate distribution between Gaussian and Cauchy distributions}
\author{Tong Liu$^{1}$, Ping Zhang$^{2}$, Wu-Sheng Dai$^{1,3}$\thanks{Email:
daiwusheng@tju.edu.cn} and Mi Xie$^{1,3}$\thanks{Email: xiemi@tju.edu.cn}\\{\footnotesize 1) Department of Physics, Tianjin University, Tianjin 300072,
P. R. China }\\{\footnotesize 2) Department of Finance, School of Economics, Nankai
University, Tianjin 300072, P. R. China}\\{\footnotesize 3) LiuHui Center for Applied Mathematics, Nankai University \&
Tianjin University, Tianjin 300072, P. R. China}}
\date{}
\maketitle

\begin{abstract}
In this paper, we construct an intermediate distribution linking the Gaussian
and the Cauchy distribution. We provide the probability density function and
the corresponding characteristic function of the intermediate distribution.
Because many kinds of distributions have no moment, we introduce weighted
moments. Specifically, we consider weighted moments under two types of
weighted functions: the cut-off function and the exponential function. Through
these two types of weighted functions, we can obtain weighted moments for
almost all distributions. We consider an application of the probability
density function of the intermediate distribution on the spectral line
broadening in laser theory. Moreover, we utilize the intermediate distribution
to the problem of the stock market return in quantitative finance.

\end{abstract}

\vskip 1cm


\section{Introduction}

In statistics, the Gaussian distribution is a standard distribution satisfying
the central limit theorem; the Cauchy distribution, also called the Lorentzian
distribution, however, typically fails to be tamed by the central limit
theorem \cite{GK}. In the meantime, these two kinds of distributions both
belong to infinitely divisible distributions \cite{Samuels,Novikov,Infinitely
Divisible}. They have some similarities, such as unimodality, symmetry and the
same domain of definition. Such similar properties are so attractive that
stimulate one to find an intermediate distribution to link up the Gaussian and
the Cauchy distribution. The Student-t distribution \cite{Preda} is the first
distribution that realizes the purpose linking the two distributions.
Nevertheless£¬the Student-t distribution is not the unique one. In this paper,
we provide another kind of distribution linking the Gaussian and the Cauchy distribution.

First, we introduce an intermediate distribution which can reduce to the
Gaussian and the Cauchy distribution at some certain values of a parameter.
Second, instead of conventional moments, we consider the weighted moment of a
distribution. Many distributions have no moments since the moment of such
distributions diverges. To seek moments, a weighted moment is introduced. By
the weighted moment, rather than the conventional moment, one can calculate
the moment in almost all kinds of distributions, such as Gaussian, Cauchy, and
Student-t distributions and the distribution provided in the present paper.
Third, we give two applications of the intermediate distribution: one is in
laser theory and the other is in quantitative finance.

Specifically, as is well known, each probability density function (PDF)
corresponds a characteristic function uniquely through a Fourier
transformation \cite{reichl,SW,Probability Analytic View}, which enlightens
one to construct a PDF starting from a characteristic function. From this
viewpoint, we provide a distribution that bridges the gap between the Gaussian
and the Cauchy distribution by analyzing their characteristic functions in
this paper.

Next, we introduce a weighted moment, because the conventional moment fails to
converge in many distributions \cite{reichl}, e.g., the Cauchy distribution
does not have conventional moments. Nevertheless, the moment is one of the
most important concepts in statistics. Many properties are reflected by each
order moment, such as deviations and variances. If a distribution does not
have moments, it makes difficulties to discuss the properties described by the
moment. Therefore, we consider the weighted moment to overcome such a defect.
We, specifically, discuss two types of weighted moments in this paper.

When considering complex systems, the concept of generalized distribution is
widely discussed. An important generalization of the Gaussian distribution is
the Tsallis distribution \cite{TMP,TRPB,RST} (the Student-t distribution
mentioned above is a special case of the Tsallis distribution). Recent
researches show that a dissipative optical lattice system will display the
Tsallis distribution \cite{DBR}. In quantitative physics, it is shown that one
can construct stock price models based on the Tsallis distribution \cite{Bor}.
In addition, some other kinds of generalized statistics and the applications
are also discussed in the literature \cite{SB1,SB2}.

Furthermore, we suggest an application of the intermediate distribution on
spectral line broadening in laser theory. The homogeneous broadening due to
collisions and spontaneous emissions is approximated by the Cauchy
distribution, and the inhomogeneous broadening due to Doppler effects is
approximated by the Gaussian distribution \cite{Principles of Lasers}. Since
the two kinds of broadenings exist simultaneously, we suggest to use the
intermediate distribution to describe the spectral line broadening. In
general, a realistic system always cannot be exactly described by an ideal
theoretical model. Therefore, some intermediate models are suggested to
describe complex systems. For example, it is found that a magnetic system
obeys a kind of intermediate statistics rather than Bose-Einstein statistics
\cite{Spinwave}. Moreover, some composite particles composed of some fermions
may behave like bosons and obey Bose-Einstein statistics when they are far
from each other; when they come closer together, however, the fermions in the
composite particles can feel each other, and the statistics of the composite
particles will deviate from exact Bose-Einstein statistics \cite{ISA}. Such a
composite-particle system will obey a certain kind of intermediate statistics.
For this reason, some intermediate theories are considered
\cite{Gentile,GentileS,GS,SU2}.

At last, we apply the intermediate distribution to describe the stock market
return in quantitative finance. The stock market return is the return that we
obtain from stock market by buying and selling stocks or get dividends by the
company whose stock you hold. Stock market returns include two parts: capital
gain and dividend income. In the Black-Scholes stock option pricing model
\cite{Pricing of Options} and the capital asset pricing model (CAPM)
\cite{market equilibrium,valuation,Equilibrium}, the stock market price is
assumed to follow lognormal distribution, which means that stock market
returns follow the Gaussian distribution. Nevertheless, in real world, stock
market returns disobey the Gaussian distribution. The characteristics of stock
market return distribution is sharp peak and fat tails. The intermediate
distribution can fit the stock market return better than the Gaussian
distribution. Moreover, in the fitting results, we also compare the
intermediate distribution with the q-Gaussian distribution, which generalizes
the Gaussian distribution in statistics \cite{New}. The distribution of the
stock market return plays a fundamental role in pricing theory. Based on the
intermediate distribution, we can price the stock price and the option in line
with the realistic.

The structure of this paper is as follows. In section 2, an intermediate
distribution linking the Gaussian and the Cauchy distribution is constructed.
In section 3, the weighted moment and some properties of the moment are
discussed. In section 4, a function is introduced to link the intermediate and
the Gaussian distribution smoothly. In section 5, an application of the
intermediate distribution to spectral line broadening in laser theory is
discussed. In section 6, the intermediate distribution is utilized to fit the
probability density function with respect to the stock market return. The
conclusion and outlook are given in Section 7.

\section{The Construction of the intermediate distribution}

In this section, we introduce an intermediate distribution. The PDF of the
intermediate distribution is
\begin{align}
p\left(  x,\mu,\sigma,\nu\right)   &  =\frac{1}{\nu\pi\exp\left(  \sigma
^{2}-\sigma^{2}/\nu^{2}\right)  }\int_{0}^{1}\cos\left(  \frac{x-\mu}{\nu}\ln
t\right)  t^{\sigma^{2}/\nu^{2}-1}\exp\left[  \left(  \sigma^{2}-\frac
{\sigma^{2}}{\nu^{2}}\right)  t\right]  dt,\nonumber\\
&  \left(  \nu>0,\text{ }\sigma>0\right)  \label{the nu distribution pdf}%
\end{align}
and the corresponding characteristic function is%
\begin{equation}
f_{\nu}\left(  k\right)  =\exp\left\{  ik\mu-\frac{\sigma^{2}}{\nu}\left\vert
k\right\vert +\sigma^{2}\left(  1-\frac{1}{\nu}\right)  \left[  \exp\left(
-\nu\left\vert k\right\vert \right)  -1\right]  \right\}  \text{.}
\label{the nu distribution cf}%
\end{equation}
Such a distribution will recover the Gaussian and the Cauchy distribution when
$\nu\rightarrow0$ and $\nu\rightarrow1$, respectively.

In the following, we deduce the PDF and the corresponding characteristic
function in detail.

\subsection{The characteristic function of the intermediate distribution}

At the earliest, the unique general formula for any characteristic function of
infinitely divisible distribution with finite variance was found by
Kolmogorov. Levy and Khintchine generalized Kolmogorov's result and discovered
the famous Levy-Khintchine formula \cite{Infinitely
Divisible,reichl,Probability Analytic View,Levy Processes}. The
Levy-Khintchine formula is also valid to all infinitely divisible distributions.

According to the Levy-Khintchine formula, all characteristic functions of
infinitely divisible distributions can be represented by \cite{Infinitely
Divisible,reichl,Probability Analytic View,Levy Processes}%
\begin{equation}
f\left(  k\right)  =\exp\left[  ik\mu+\int_{-\infty}^{+\infty}\left(
e^{iky}-1-\frac{iky}{1+y^{2}}\right)  \frac{1+y^{2}}{y^{2}}dG\left(  y\right)
\right]  , \label{1}%
\end{equation}
where $\mu$ is a real constant and $G\left(  y\right)  $ is a real, bounded,
and nondecreasing function of $y$. Different $G\left(  y\right)  $'s generate
different distributions.

What we want to do is to introduce a function linking Gaussian and Cauchy
distributions. We construct the function $G\left(  y,\nu\right)  $ as
\begin{equation}
G\left(  y,\nu\right)  =\frac{\sigma^{2}}{\pi}\arctan\frac{y}{\nu}%
+\frac{\sigma^{2}}{2},\text{ }(\nu>0,\text{ }\sigma>0). \label{2}%
\end{equation}
Obviously£¬$G\left(  y,\nu\right)  $ is a real, bounded, and nondecreasing
function of $y$ from $-\infty$ to $+\infty$. Because $G\left(  y,\nu\right)  $
satisfies the requirements of the Levy-Khintchine formula, it corresponds to a
characteristic function of a certain distribution. The function $G\left(
y,\nu\right)  $ recovers the characteristic functions of Gaussian and Cauchy
distributions when $\nu\rightarrow0$ and $\nu\rightarrow1 $, respectively:
\begin{equation}
G\left(  y,0\right)  =G_{Gauss}\left(  y\right)  =\sigma^{2}\Theta\left(
y\right)  , \label{3}%
\end{equation}%
\begin{equation}
G\left(  y,1\right)  =G_{Cauchy}\left(  y\right)  =\frac{\sigma^{2}}{\pi
}\arctan y+\frac{\sigma^{2}}{2},
\end{equation}
where $\Theta\left(  y\right)  $ is the step function. In this paper, the
intermediate distribution is generated from $G\left(  y,\nu\right)  $ in Eq.
(\ref{2}).

For simplicity£¬introduce%
\begin{equation}
H\left(  y\right)  =\frac{\partial G}{\partial y}=\frac{\sigma^{2}\nu}%
{\pi\left(  y^{2}+\nu^{2}\right)  }. \label{7}%
\end{equation}
Inserting $H\left(  y\right)  $ into the Levy-Khintchine formula (\ref{1}), we
have
\begin{align}
f_{\nu}\left(  k\right)   &  =\exp\left[  ik\mu+\int_{-\infty}^{+\infty
}e^{iky}\frac{1+y^{2}}{y^{2}}H\left(  y\right)  dy\right. \nonumber\\
&  \left.  -\int_{-\infty}^{+\infty}\frac{1+y^{2}}{y^{2}}H\left(  y\right)
dy-ik\int_{-\infty}^{+\infty}\frac{H\left(  y\right)  }{y}dy\right]  .
\label{8}%
\end{align}

Here
\begin{equation}
\mathcal{P}\int_{-\infty}^{+\infty}\frac{H\left(  y\right)  }{y}dy=0,
\label{9}%
\end{equation}
where $\mathcal{P}$ represents the Cauchy principal value. Performing the
Fourier transformation, the result of the first integral in Eq. (\ref{8})
becomes%
\begin{equation}
\int_{-\infty}^{+\infty}e^{iky}\frac{1+y^{2}}{y^{2}}H\left(  y\right)
dy=-\frac{\sigma^{2}}{\nu}\left\vert k\right\vert +\sigma^{2}\left(
1-\frac{1}{\nu^{2}}\right)  \exp\left(  -\nu\left\vert k\right\vert \right)  ,
\label{10}%
\end{equation}
and the second integral is%
\begin{equation}
\int_{-\infty}^{+\infty}\frac{1+y^{2}}{y^{2}}H\left(  y\right)  dy=\sigma
^{2}\left(  1-\frac{1}{\nu^{2}}\right)  . \label{11}%
\end{equation}
Finally£¬the characteristic function of the intermediate distribution
generated by Eq. (\ref{2}) is shown as Eq. (\ref{the nu distribution cf}).

\subsection{The PDF of the intermediate distribution}

Now, we deduce the PDF of the intermediate distribution corresponding to the
characteristic function given by Eq. (\ref{the nu distribution cf}).

Facilitating the inverse Fourier transformation, we rewrite $f_{\nu}\left(
k\right)  $ as%
\begin{align}
f_{\nu}\left(  k\right)   &  =\exp\left\{  ik\mu-\frac{\sigma^{2}}{\nu
}\left\vert k\right\vert +\sigma^{2}\left(  1-\frac{1}{\nu^{2}}\right)
\left[  \exp\left(  -\nu\left\vert k\right\vert \right)  -1\right]  \right\}
\nonumber\\
&  =\exp\left[  ik\mu-\sigma^{2}\left(  1-\frac{1}{\nu^{2}}\right)  \right]
\nonumber\\
&  \times\left\{  \exp\left[  \frac{\sigma^{2}}{\nu}k+\sigma^{2}\left(
1-\frac{1}{\nu^{2}}\right)  \exp\left(  \nu k\right)  \right]  \Theta\left(
-k\right)  \right. \nonumber\\
&  \left.  +\exp\left[  -\frac{\sigma^{2}}{\nu}k+\sigma^{2}\left(  1-\frac
{1}{\nu^{2}}\right)  \exp\left(  -\nu k\right)  \right]  \Theta\left(
k\right)  \right\}  .
\end{align}
Then, performing the inverse Fourier transformation to $f_{\nu}\left(
k\right)  $ gives%
\begin{align}
&  p\left(  x,\mu,\sigma,\nu\right)  =\frac{1}{\sqrt{2\pi}}\mathcal{F}%
^{-1}\left[  f_{\nu}\left(  k\right)  ,k,x\right] \nonumber\\
&  =\frac{1}{\nu\pi}\exp\left[  \sigma^{2}\left(  \frac{1}{\nu^{2}}-1\right)
\right]  \left[  \sigma^{2}\left(  \frac{1}{\nu^{2}}-1\right)  \right]
^{-\sigma^{2}/\nu^{2}}\nonumber\\
&  \times\frac{1}{2}\left\{  \left[  \sigma^{2}\left(  \frac{1}{\nu^{2}%
}-1\right)  \right]  ^{-i\left(  x-\mu\right)  /\nu}\Gamma\left[  i\frac
{x-\mu}{\nu}+\frac{\sigma^{2}}{\nu^{2}},0,\sigma^{2}\left(  \frac{1}{\nu^{2}%
}-1\right)  \right]  \right. \nonumber\\
&  \left.  +\left[  \sigma^{2}\left(  \frac{1}{\nu^{2}}-1\right)  \right]
^{i\left(  x-\mu\right)  /\nu}\Gamma\left[  -i\frac{x-\mu}{\nu}+\frac
{\sigma^{2}}{\nu^{2}},0,\sigma^{2}\left(  \frac{1}{\nu^{2}}-1\right)  \right]
\right\}  ,
\end{align}
where $\Gamma\left(  \alpha,z_{0},z_{1}\right)  =\int_{z_{0}}^{z_{1}}%
t^{\alpha-1}e^{-t}dt$ is the generalized incomplete gamma function. Here
$p\left(  x,\mu,\sigma,\nu\right)  $ is a real function of $x$, because the
two parts in the braces are complex conjugate and can be rewritten as a
manifest real form as Eq. (\ref{the nu distribution pdf}).

When $\nu\rightarrow0$ and $\nu\rightarrow1$, the intermediate distribution
$p\left(  x,\mu,\sigma,\nu\right)  $ given by Eq. (\ref{the nu distribution
pdf}) reduces to the Gaussian distribution%
\begin{equation}
p\left(  x,\mu,\sigma,0\right)  =\frac{1}{\sqrt{2\pi}\sigma}\exp\left[
-\frac{\left(  x-\mu\right)  ^{2}}{2\sigma^{2}}\right]  \label{Gaussian pdf}%
\end{equation}
and the Cauchy distribution
\begin{equation}
p\left(  x,\mu,\sigma,1\right)  =\frac{\sigma^{2}}{\pi}\frac{1}{\sigma
^{4}+\left(  x-\mu\right)  ^{2}},
\end{equation}
respectively.

\section{The weighted moment}

In this section, we introduce weighted moments and discuss their behaviors.

The reason why we consider the weighted moment is that the conventional moment
diverges in many distributions, such as the Cauchy distribution. To obtain and
compare the moment between different distributions, we introduce the weighted
function to seek a convergent moment. A parameter is used to adjust the
weighted function. When the parameter takes a certain value, the weighted
function reduces to $1$ and coincides with the conventional one. Specifically,
we discuss the weighted moment with the cut-off weighted function and the
exponential weighted function.

The moment is one of the essential concepts in statistics. It is because that
the even order moment, especially the second order moment, reflects the
deviation of a quantity that is estimated. It is inconvenient for us to
analyze the deviation of this quantity that a distribution does not have the
even order moments. Moreover, the mean value of a quantity can be expanded as
a series of the moments since the quantity can be technically expanded as a
power series.

\subsection{The moment and the weighted moment}

In this section, we introduce the weighted moment.

The central moment in statistics is defined as%
\begin{equation}
\left\langle x^{m}\right\rangle =\int_{-\infty}^{+\infty}\left(  x-\mu\right)
^{m}p\left(  x-\mu\right)  dx=\int_{-\infty}^{+\infty}x^{m}p\left(  x\right)
dx\text{ \ \ }\left(  m\geq0\right)  . \label{27}%
\end{equation}
Note that one also considers another kind of moment, the origin moment
$\left\langle x^{m}\right\rangle ^{origin}=\int_{-\infty}^{+\infty}%
x^{m}p\left(  x-\mu\right)  dx$ with $m\geq0$, in statistics. We can easily
find that the value of the origin moment depends on the central site $x=\mu$
of the distribution, while the central moment is independent on the site
$x=\mu$. Thus, the central moment is more natural than the origin moment. We
will only discuss the central moment in this paper.

If $p\left(  x\right)  $ is an even function, the Cauchy principal value of
the odd order moments vanishes%
\begin{equation}
\left\langle x^{2n+1}\right\rangle =\mathcal{P}\int_{-\infty}^{+\infty
}x^{2n+1}p\left(  x\right)  dx=0,\text{ }\left(  n=0,1,2,\cdots\right)  .
\end{equation}
The even order moment reads%
\begin{equation}
\left\langle x^{2n}\right\rangle =\int_{-\infty}^{+\infty}x^{2n}p\left(
x\right)  dx,\text{ }\left(  n=0,1,2,\cdots\right)  .
\end{equation}

The Gaussian distribution has all even order central moments:
\begin{equation}
\left\langle x^{2n}\right\rangle =\int_{-\infty}^{+\infty}dx\frac{x^{2n}%
}{\sqrt{2\pi}\sigma}\exp\left(  -\frac{x^{2}}{2\sigma^{2}}\right)  =\frac
{1}{\sqrt{\pi}}2^{n}\sigma^{2n}\Gamma\left(  n+\frac{1}{2}\right)  ,
\label{the Gaussian moment}%
\end{equation}
where $\Gamma\left(  z\right)  $ is the Euler gamma function. As is well
known, the Cauchy distribution, however, does not has central moments. The
intermediate distribution given by Eq. (\ref{the nu distribution pdf}), like
the Cauchy distribution, has no central moments. The $2n$-th moment of Eq.
(\ref{the nu distribution pdf})
\begin{equation}
\left\langle x^{2n}\right\rangle =\frac{1}{\nu\pi\exp\left(  \sigma^{2}%
-\frac{\sigma^{2}}{\nu^{2}}\right)  }\int_{0}^{1}dtt^{\sigma^{2}/\nu^{2}%
-1}\exp\left[  \left(  \sigma^{2}-\frac{\sigma^{2}}{\nu^{2}}\right)  t\right]
\int_{-\infty}^{+\infty}dxx^{2n}\cos\left(  x\frac{\ln t}{\nu}\right)
\end{equation}
diverges, because the integral $\int_{-\infty}^{+\infty}dxx^{m}\cos\left(
x\frac{\ln t}{\nu}\right)  $ does not converge on $\left(  -\infty
,+\infty\right)  $.

In order to find a quantity to play the role of moments to the distribution
which does not have moments, we define a weighted moment as
\begin{equation}
\left\langle x^{m}\right\rangle _{w}=\int_{-\infty}^{+\infty}x^{m}w\left(
x,\lambda\right)  p\left(  x,0,\sigma,\nu\right)  dx, \label{28}%
\end{equation}
by introducing a symmetrical weighted function $w\left(  x,\lambda\right)  $.
Since $w\left(  x,\lambda\right)  $ and $p\left(  x,0,\sigma,\nu\right)  $ are
symmetrical, the Cauchy principal value of the odd order moments vanishes,
and, then, only the even order moment
\begin{equation}
\left\langle x^{2n}\right\rangle _{w}=\int_{-\infty}^{+\infty}x^{2n}w\left(
x,\lambda\right)  p\left(  x,0,\sigma,\nu\right)  dx,\text{ \ \ }\left(
n=0,1,2,\cdots\right)
\end{equation}
needs to be considered.

In the following, we consider two weighted functions.

\subsection{The weighted moment with the cut-off weighted function $w\left(
x,a\right)  =\Theta\left(  x+a\right)  -\Theta\left(  x-a\right)  $}

In this section, we consider a cut-off weighted function,
\begin{equation}
w\left(  x,a\right)  =\Theta\left(  x+a\right)  -\Theta\left(  x-a\right)  .
\label{cut-off weight}%
\end{equation}
The even order weighted moment with the weighted function (\ref{cut-off
weight}) is\ then
\begin{align}
\left\langle x^{2n}\right\rangle _{\mathrm{cut-off}}  &  =\int_{-\infty
}^{+\infty}x^{2n}\left[  \Theta\left(  x+a\right)  -\Theta\left(  x-a\right)
\right]  p\left(  x,0,\sigma,x\right) \nonumber\\
&  =\int_{-a}^{+a}x^{2n}p\left(  x,0,\sigma,\nu\right)  dx,\text{ \ \ }\left(
n=0,1,2,\cdots\right)  ,
\end{align}
which reduces to the moment Eq. (\ref{27}) when $a\rightarrow\infty$.

Finally, we can obtain the cut-off weighted moments, $\left\langle
x^{2n}\right\rangle _{\mathrm{cut-off}}$.

\textit{The weighted moment of the intermediate distribution:}
\begin{align}
\left\langle x^{2n}\right\rangle _{\mathrm{cut-off}}  &  =\int_{-a}^{+a}%
x^{2n}p\left(  x,0,\sigma,\nu\right)  dx\nonumber\\
&  =\frac{1}{\nu\pi\exp\left(  \sigma^{2}-\sigma^{2}/\nu^{2}\right)  }%
\frac{a^{2n+1}}{2n+1}\nonumber\\
&  \times\int_{0}^{1}dtt^{\sigma^{2}/\nu^{2}-1}\exp\left[  \left(  \sigma
^{2}-\frac{\sigma^{2}}{\nu^{2}}\right)  t\right]  \text{ }_{1}F_{2}\left[
n+\frac{1}{2};\frac{1}{2},n+\frac{3}{2};-\frac{1}{4}\left(  \frac{\ln t}{\nu
}\right)  ^{2}a^{2}\right] \nonumber\\
&  =\frac{\nu}{\pi\sigma^{2}\exp\left(  \sigma^{2}-\sigma^{2}/\nu^{2}\right)
}a^{2n+1}\sum_{k=0}^{\infty}\frac{\left(  -1\right)  ^{k}}{\left(
2k+2n+1\right)  \left(  2k\right)  !}\left(  \frac{\nu}{\sigma^{2}}\right)
^{2k}\nonumber\\
&  \times\text{ }_{2k+1}F_{2k+1}\left[  \left\{  \frac{\sigma^{2}}{\nu^{2}%
},\cdots,\frac{\sigma^{2}}{\nu^{2}}\right\}  ;\left\{  1+\frac{\sigma^{2}}%
{\nu^{2}},\cdots,1+\frac{\sigma^{2}}{\nu^{2}}\right\}  ;\sigma^{2}%
-\frac{\sigma^{2}}{\nu^{2}}\right]  a^{2k}, \label{DMM}%
\end{align}
Here, $\Gamma\left(  \alpha\right)  =\Gamma\left(  \alpha,0,\infty\right)  $,
$\Gamma\left(  \alpha,z_{0}\right)  =\Gamma\left(  \alpha,z_{0},\infty\right)
$, $_{p}F_{q}\left[  \left\{  a_{1},\cdots,a_{p}\right\}  ;\left\{
b_{1},\cdots,b_{q}\right\}  ;z\right]  $ is the generalized hypergeometric
function, and the integral
\begin{align}
&  \int_{0}^{1}t^{\sigma^{2}/\nu^{2}-1}\exp\left[  \left(  \sigma^{2}%
-\frac{\sigma^{2}}{\nu^{2}}\right)  t\right]  \left(  \ln t\right)
^{2k}dt\nonumber\\
&  =\left(  \frac{\nu^{2}}{\sigma^{2}}\right)  ^{2k+1}\text{ }_{2k+1}%
F_{2k+1}\left[  \left\{  \frac{\sigma^{2}}{\nu^{2}},\cdots,\frac{\sigma^{2}%
}{\nu^{2}}\right\}  ;\left\{  1+\frac{\sigma^{2}}{\nu^{2}},\cdots
,1+\frac{\sigma^{2}}{\nu^{2}}\right\}  ;\sigma^{2}-\frac{\sigma^{2}}{\nu^{2}%
}\right]
\end{align}
is used.

\textit{The weighted moment of the Cauchy distribution:}
\begin{align}
\left\langle x^{2n}\right\rangle _{\mathrm{cut-off}}  &  =\int_{-a}^{+a}%
x^{2n}p\left(  x,0,\sigma,1\right)  dx\nonumber\\
&  =\left(  -1\right)  ^{n}\sigma^{4n}+\frac{2\sigma^{2}a^{2n-1}}{\pi\left(
2n-1\right)  }\text{ }_{2}F_{1}\left[  \left\{  1,\frac{1}{2}-n\right\}
;\left\{  \frac{3}{2}-n\right\}  ;-\frac{\sigma^{4}}{a^{2}}\right] \nonumber\\
&  =\left(  -1\right)  ^{n}\sigma^{4n}+\frac{2a^{2n+1}}{\pi\sigma^{2}}%
\sum_{k=1}^{\infty}\frac{\left(  -1\right)  ^{k}\sigma^{4k}}{2k-2n-1}a^{-2k}.
\end{align}

\textit{The weighted moment of the Gaussian distribution:}
\begin{equation}
\left\langle x^{2n}\right\rangle _{\mathrm{cut-off}}=\int_{-a}^{+a}%
x^{2n}p\left(  x,0,\sigma,0\right)  dx=\frac{2^{n}\sigma^{2n}}{\sqrt{\pi}%
}\left[  \Gamma\left(  n+\frac{1}{2}\right)  -\Gamma\left(  n+\frac{1}%
{2},\frac{a^{2}}{2\sigma^{2}}\right)  \right]  . \label{DGM}%
\end{equation}
When $a\rightarrow\infty$, the cut-off weighted moment of the Gaussian
distribution reduces to the moment Eq. (\ref{the Gaussian moment}).

\subsection{The weighted moment with the exponential weighted function
$w\left(  x,\alpha\right)  =e^{-\frac{1}{2}\alpha^{2}x^{2}}$}

In this section, we consider the exponential weight function defined as
\begin{equation}
w\left(  x,\alpha\right)  =e^{-\frac{1}{2}\alpha^{2}x^{2}}.
\label{exponential weight}%
\end{equation}
The even order weighted moment with the weighted function (\ref{exponential
weight}) is then
\begin{equation}
\left\langle x^{2n}\right\rangle _{\exp}=\int_{-\infty}^{+\infty}%
x^{2n}e^{-\frac{1}{2}\alpha^{2}x^{2}}p\left(  x,0,\sigma,\nu\right)  dx,\text{
\ \ }\left(  n=0,1,2,\cdots\right)  ,
\end{equation}
which reduces to the moment Eq. (\ref{27}) when $\alpha\rightarrow0$.

Finally, we obtain the exponential weighted moments, $\left\langle
x^{2n}\right\rangle _{\exp}$.

\textit{The weighted moment of the intermediate distribution:}
\begin{align}
\left\langle x^{2n}\right\rangle _{\exp}  &  =\int_{-\infty}^{+\infty}%
x^{2n}e^{-\frac{1}{2}\alpha^{2}x^{2}}p\left(  x,0,\sigma,\nu\right)
dx,\nonumber\\
&  =\frac{2^{n+\frac{1}{2}}\nu}{\pi\sigma^{2}e^{\sigma^{2}-\frac{\sigma^{2}%
}{\nu^{2}}}}\alpha^{-2n-1}\sum_{k=0}^{\infty}\left(  \frac{-2\nu^{2}}%
{\sigma^{4}}\right)  ^{k}\Gamma\left(  k+n+\frac{1}{2}\right) \nonumber\\
&  \times\text{ }_{2k+1}F_{2k+1}\left[  \left\{  \frac{\sigma^{2}}{\nu^{2}%
},\cdots,\frac{\sigma^{2}}{\nu^{2}}\right\}  ;\left\{  1+\frac{\sigma^{2}}%
{\nu^{2}},\cdots,1+\frac{\sigma^{2}}{\nu^{2}}\right\}  ;\sigma^{2}%
-\frac{\sigma^{2}}{\nu^{2}}\right]  \alpha^{-2k}. \label{31}%
\end{align}

\textit{The weighted moment of the Cauchy distribution:}
\begin{align}
\left\langle x^{2n}\right\rangle _{\exp}  &  =\int_{-\infty}^{+\infty}%
x^{2n}e^{-\frac{1}{2}\alpha^{2}x^{2}}p\left(  x,0,\sigma,1\right)
dx\nonumber\\
&  =\frac{1}{\pi}\sigma^{4n}\Gamma\left(  n+\frac{1}{2}\right)  e^{\frac{1}%
{2}\sigma^{4}\alpha^{2}}\Gamma\left(  \frac{1}{2}-n,\frac{1}{2}\sigma
^{2}\alpha^{4}\right) \nonumber\\
&  =\left(  -1\right)  ^{n}\sigma^{4n}e^{\frac{1}{2}\sigma^{2}\alpha^{4}%
}+\frac{\sigma^{2}}{2\pi}2^{n+\frac{1}{2}}\Gamma\left(  n-\frac{1}{2}\right)
\alpha^{-2n+1}\sum_{k=0}^{\infty}\frac{\Gamma\left(  k-n+\frac{3}{2}\right)
}{\Gamma\left(  -n+\frac{3}{2}\right)  }2^{-k}\sigma^{4k}\alpha^{2k}.
\label{30}%
\end{align}

\textit{The weighted moment of the Gaussian distribution:}
\begin{equation}
\left\langle x^{2n}\right\rangle _{\exp}=\int_{-\infty}^{+\infty}%
x^{2n}e^{-\frac{1}{2}\alpha^{2}x^{2}}p\left(  x,0,\sigma,0\right)  dx=\frac
{1}{\sqrt{\pi}\sigma}2^{n}\Gamma\left(  n+\frac{1}{2}\right)  \left(
\alpha^{2}+\frac{1}{\sigma^{2}}\right)  ^{-n-\frac{1}{2}}. \label{29}%
\end{equation}
When $\alpha\rightarrow0$, the weighted moment of the Gaussian distribution
reduces to the moment Eq. (\ref{the Gaussian moment}).

\section{An alternative construction of the intermediate distribution}

In this section, we suggest another construction of the PDF of the
intermediate distribution, which links the intermediate distribution and the
Gaussian distribution smoothly.

In the PDF of the intermediate distribution, Eq. (\ref{the nu distribution
pdf}), the points $\nu=0$ is a removable singularity. According to the
analysis in the above section, the behavior of the moment depends on the value
of the parameter $\nu$. The distribution have no moments unless $\nu
\rightarrow0$ so that the intermediate distribution, Eq. (\ref{the nu
distribution pdf}), reduces to the Gaussian distribution. What we want to do
is to seek a certain way that when the parameter $\nu$ runs smoothly from a
constant $\nu_{0}$ to $0$, the distribution travels smoothly from the
intermediate distribution to the Gaussian distribution. Especially, when
$\nu_{0}=1$, the distribution will travel smoothly from the Cauchy
distribution to the Gaussian distribution.

Specifically, we insert a smooth function $h\left(  t,\nu_{0},a,b\right)  $
into the PDF of the intermediate distribution, Eq. (\ref{the nu distribution
pdf}), and then have%
\begin{align}
&  p\left(  x,\mu,\sigma,\nu_{0},t,a,b\right) \nonumber\\
&  =\frac{1}{\nu\pi\exp\left[  \sigma^{2}-\sigma^{2}/h^{2}\left(  t,\nu
_{0},a,b\right)  \right]  }\nonumber\\
&  \times\int_{0}^{1}\cos\left[  \frac{x-\mu}{h\left(  t,\nu_{0},a,b\right)
}\ln\xi\right]  \xi^{\sigma^{2}/h^{2}\left(  t,\nu_{0},a,b\right)  -1}%
\exp\left[  \left(  \sigma^{2}-\frac{\sigma^{2}}{h^{2}\left(  t,\nu
_{0},a,b\right)  }\right)  \xi\right]  d\xi, \label{p(x,h(t,a,b))}%
\end{align}
where the function $h\left(  t,\nu_{0},a,b\right)  $ is \cite{S. S. Chern,S.
P. Novikov}
\begin{equation}
h\left(  t,\nu_{0},a,b\right)  =\left\{
\begin{array}
[c]{l}%
\nu_{0},\text{ \ \ }t\in\left(  -\infty,a\right]  ,\\
\frac{\nu_{0}}{\lambda_{0}}\int_{t}^{b}s\left(  \xi,a,b\right)  d\xi,\text{
\ \ }t\in\left(  a,b\right)  ,\\
0,\text{ \ \ }t\in\left[  b,+\infty\right)  ,
\end{array}
\right.  \label{20}%
\end{equation}
with $\lambda_{0}=\int_{a}^{b}s\left(  \xi,a,b\right)  d\xi$, and
\begin{equation}
s\left(  t,a,b\right)  =\left\{
\begin{array}
[c]{l}%
\exp\left(  \frac{1}{t-b}-\frac{1}{t-a}\right)  ,\text{ \ \ }t\in\left(
a,b\right)  ,\\
0,\text{ \ \ }t\in\left(  -\infty,a\right]  \cup\left[  b,+\infty\right)  .
\end{array}
\right.  \label{19}%
\end{equation}
The figures of $h\left(  t,\nu_{0},a,b\right)  $ and $s\left(  t,a,b\right)  $
are shown in Figure 1.

\begin{figure}[ptb]
\begin{center}
\includegraphics[
width=10cm ]{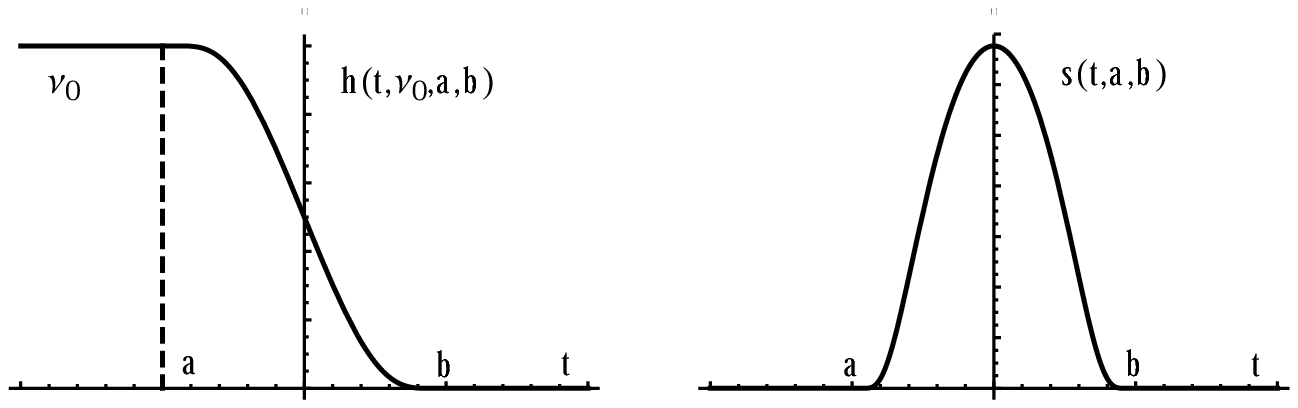}
\end{center}
\caption{The function $h\left(  t,\nu_{0},a,b\right)  $ Eq. (\ref{20}) and
$s\left(  t,a,b\right)  $ Eq. (\ref{19})}%
\end{figure}

When $t\geq b$, $h\left(  t,\nu_{0},a,b\right)  =0$, and the PDF Eq.
(\ref{p(x,h(t,a,b))}) reduces to the Gaussian PDF. When $\nu_{0}=1$ and $t\leq
a$, $h\left(  t,1,a,b\right)  =1$, the PDF Eq. (\ref{p(x,h(t,a,b))}) reduces
to the Cauchy PDF. When $t$ runs from $a$ to $b$, the distribution travels
from the intermediate distribution to the Gaussian distribution smoothly.
Especially, when $\nu_{0}=1$, the distribution travels from the Cauchy
distribution to the Gaussian distribution.

\section{Application to the problem of the spectral line broadening in laser
theory}

In this section, we apply the intermediate distribution introduced in the
present paper to the spectral line broadening in laser theory.

In laser theory, the spectral line broadening is a phenomenon due to photons
emitted or absorbed in a narrow frequency range. The spectral line broadening
of laser can be approximated by a distribution. The Cauchy distribution and
Gaussian distribution are often used to describe the broadening: the Cauchy
distribution is used to describe the homogeneous broadening; the Gaussian
distribution is used to describe the inhomogeneous broadening. Nevertheless,
the homogeneous and inhomogeneous broadening exist simultaneously. That is to
say, in a system, the broadening is neither homogeneous nor inhomogeneous, but
between homogeneous and inhomogeneous. Therefore, we need a function to
describe the mixed broadening. In this paper, we suggest that, instead of the
Gaussian and the Cauchy distribution, one can use an intermediate distribution
to describe the spectral line broadening.

In the semiclassical approach of spectral line broadening in laser system, the
material and the light are described by quantum mechanics and the Maxwell
equations, respectively, to calculate the interaction between the medium and
the light \cite{Effect Laser Fano}. Under the electric dipole approximation,
the rate of absorption and stimulated emission of the medium between two
energy levels is \cite{Principles of Lasers}%
\begin{equation}
W=\frac{\pi}{3n^{2}\epsilon_{0}\hbar^{2}}\left\vert \mathbf{p}\right\vert
^{2}\rho g\left(  \omega-\omega_{0}\right)  , \label{23}%
\end{equation}
where $n$ is the refractive index of the material, $\epsilon_{0}$ is the
vacuum permittivity, $\mathbf{p}$ is the electric dipole moment, $\rho$ is the
energy density of the light, $\omega_{0}$ is the frequency coinciding with the
interval of two energy levels defined by $\omega_{0}=\left\vert E_{2}%
-E_{1}\right\vert /\hbar$. The function $g\left(  \omega-\omega_{0}\right)  $
in Eq. (\ref{23}) can be expressed by a probability density distribution using
to describe the broadening near the frequency $\omega_{0}$. It is clear that
the rate of absorption and stimulated emission is determined by the function
$g\left(  \omega-\omega_{0}\right)  $.

If the life time of the system at a certain energy level is infinite,
$g\left(  \omega-\omega_{0}\right)  $ reduces to $\delta\left(  \omega
-\omega_{0}\right)  $, where $\delta\left(  x\right)  $ is the Dirac delta
function. In a realistic system, the life time of an excited state is finite.
There are many kinds of factors effecting on a laser system, such as
collisions, spontaneous emission, Doppler effect, etc. These factors lead to
the broadening, and make $g\left(  \omega-\omega_{0}\right)  $ deviate from
$\delta\left(  \omega-\omega_{0}\right)  $. This can be directly measured in
experiments \cite{Laser Experiment}. We usually use full width at half maximum
$\omega_{FWHM}$ of $g\left(  \omega-\omega_{0}\right)  $ to describe the
effect of the broadening. The reciprocal of full width at half maximum
$\omega_{FWHM}$ indicates the life time of the level.

There are two kinds of mechanisms in spectral line broadening in a laser
system \cite{Handbook of Laser}. One is homogeneous which arises from
collisions and spontaneous emission, and it can be described by the Cauchy
distribution. The other is inhomogeneous broadening which arises from the
Doppler effect and the dislocation of the material. The Doppler broadening can
be described by the Gaussian distribution, but the broadening caused by the
dislocation has no proper expression. In general, homogeneous and
inhomogeneous broadenings appear in a laser system simultaneously. Therefore,
the function $g\left(  \omega-\omega_{0}\right)  $ is actually between the
Cauchy and Gaussian PDF.

\begin{figure}[ptb]
\begin{center}
\includegraphics[width=10cm]{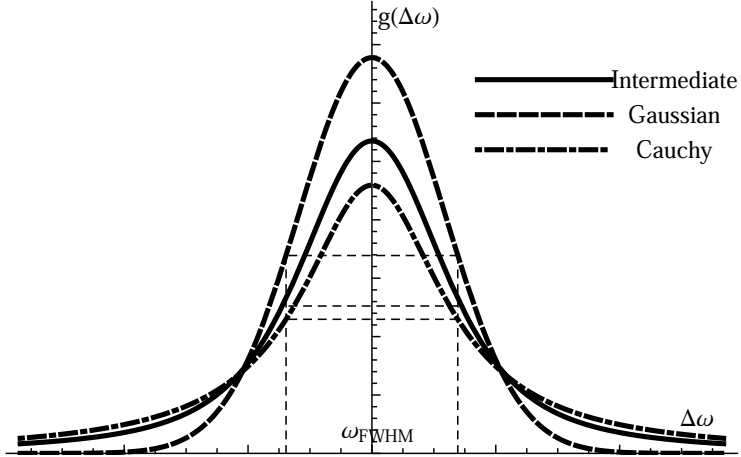}
\end{center}
\caption{The Gaussian, the Cauchy, and the intermediate type line shape
function $g\left(  \Delta\omega\right)  $ with the same full width at half
maximum.}%
\end{figure}

\begin{figure}[ptb]
\begin{center}
\includegraphics[width=10cm]{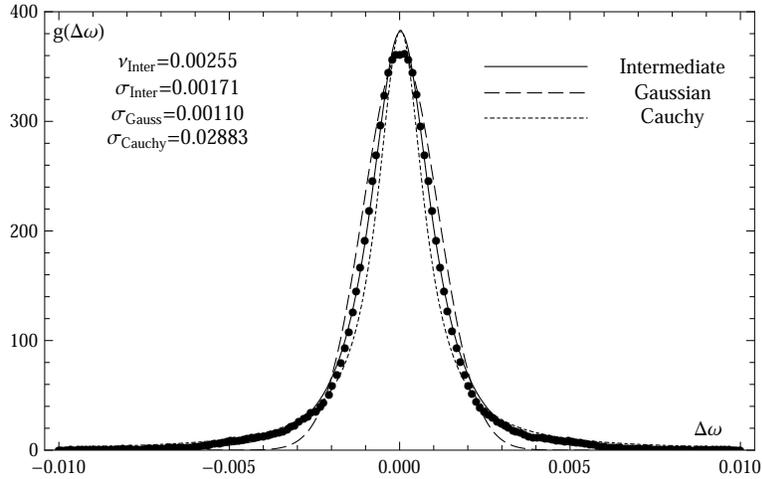}
\end{center}
\caption{The fitting of the experimental line width data with the
intermediate, the Gaussian, and the Cauchy distribution. The data are taken
from Ref. \cite{spectral data}}%
\end{figure}

Since the intermediate distribution given by Eq. (\ref{the nu distribution
pdf}) is between the Gaussian and the Cauchy distributions, we suggest that
one can use the intermediate PDF, Eq. (\ref{the nu distribution pdf}), to
describe the function $g\left(  \omega-\omega_{0}\right)  $. Specifically,
take%
\begin{equation}
g\left(  \omega-\omega_{0}\right)  =p\left(  \omega,\omega_{0},\sigma
,\nu\right)  , \label{24}%
\end{equation}
where $p\left(  x,\mu,\sigma,\nu\right)  $ is defined by the Eq. (\ref{the nu
distribution pdf}). The parameters $\sigma$ and $\nu$ play the roles of
fitting constants. The line shape of the broadenings which have the same full
width at half maximum $\omega_{FWHM}$ is shown in Figure 2.

We import a set of data of hydrogen Zeeman-Stark line broadening from Figure 2
(b) in Ref. \cite{spectral data} and fit them with the intermediate, Cauchy,
and Gaussian distribution, respectively. The data and the fitting results are
plotted in Figure 3. From the fitting results, the intermediate distribution
describes the line broadening better than the Gaussian and the Cauchy distribution.

\section{Application to stock market returns}

Gaussian distributed return is one of the most important assumptions in
quantitative finance \cite{market equilibrium,valuation,Equilibrium}. The key
reason why we make this assumption is that the Gaussian distribution is easy
to handle and we can get a lot of analytical results. The Gaussian
distribution assumption, however, has got much criticism for its failure to
describe the real world. Stock market returns in real world have sharp peak
and fat tails distribution, which means that the probability of stock market
crash is high. If we take the Gaussian distribution, the stock market crash is
almost impossible to happen. Nevertheless, the stock market crash is not far
away from us, like the Great Depression and the Black Thursday. We know that
the Gaussian distribution does not have those characteristics. Thus we deem
that the Gaussian distribution is a poor model of stock market return, and it
is inappropriate to use the Gaussian distribution to describe the stock
market. In this section, the market return is fitted by the intermediate
distribution. We use S\&P 500 price index, Dow Jones index, Nasdaq composite
index, and Nikkei 225 index. The daily stock prices are obtained from Wind
Database -- a staple financial database. All those indexes are price index,
and we compute the return from this formula%
\[
r_{t}=\frac{p_{t}-p_{t-1}}{p_{t-1}},
\]
where $r_{t}$ is the return and $p_{t}$ is the stock price.

The q-Gaussian distribution is an important distribution, which has
wide-ranging applications in various research scopes, as well as financial
physics \cite{New1,New2,New3}. The q-Gaussian distribution is derived from the
minimum Tsallis entropy \cite{New5,New6}
\begin{equation}
S_{q}\left[  p\right]  =\frac{1}{2q}\left\{  1-\int_{-\infty}^{+\infty}\left[
p\left(  x\right)  \right]  ^{2q+1}dx\right\}  ,\text{ \ \ }\left(
0<q<1\right)  ,
\end{equation}
with the constraints of normalization of probability $1=\int_{-\infty
}^{+\infty}p_{q}\left(  x\right)  dx$ and a constant variance $\sigma_{q}%
^{2}=\int_{-\infty}^{+\infty}x^{2}p_{q}\left(  x\right)  dx$. The probability
density function of the q-Gaussian distribution is
\begin{equation}
p\left(  x,\mu,\sigma,q\right)  =\frac{1}{A_{q}\sqrt{\pi}\sigma_{q}}\exp
_{q}\left[  -\frac{\left(  x-\mu\right)  ^{2}}{2\sigma_{q}^{2}}\right]  ,
\end{equation}
where $A_{q}=\sqrt{\frac{1-q}{q}}\Gamma\left(  \frac{1-q}{2q}\right)
/\Gamma\left(  \frac{1}{2q}\right)  $ and the q-exponential function is
$\exp_{q}\left(  z\right)  =\left(  1-\frac{2q}{1-q}z\right)  ^{-1/\left(
2q\right)  }$. When $q\rightarrow0$, the q-Gaussian distribution returns to
the Gaussian distribution, Eq. (\ref{Gaussian pdf}).

Making use of the intermediate, the q-Gaussian and the Gaussian distribution
to fit the data, we plot the results in Figure 4-7. We can find that the
intermediate and the q-Gaussian distribution are much better than the Gaussian
distribution. The real data have sharp peak and fat tails and the intermediate
and the q-Gaussian distribution both give good fits to this phenomenon. From
these figures we can also find that the intermediate distribution is somewhat
more suitable to describe the peak of the data of stock market returns,
especially for the Nikkei index, and the log-log plots indicate that the
q-Gaussian distribution describes the fat tails better than the intermediate distribution.

\begin{figure}[ptb]
\begin{center}
\includegraphics[width=6.5cm]{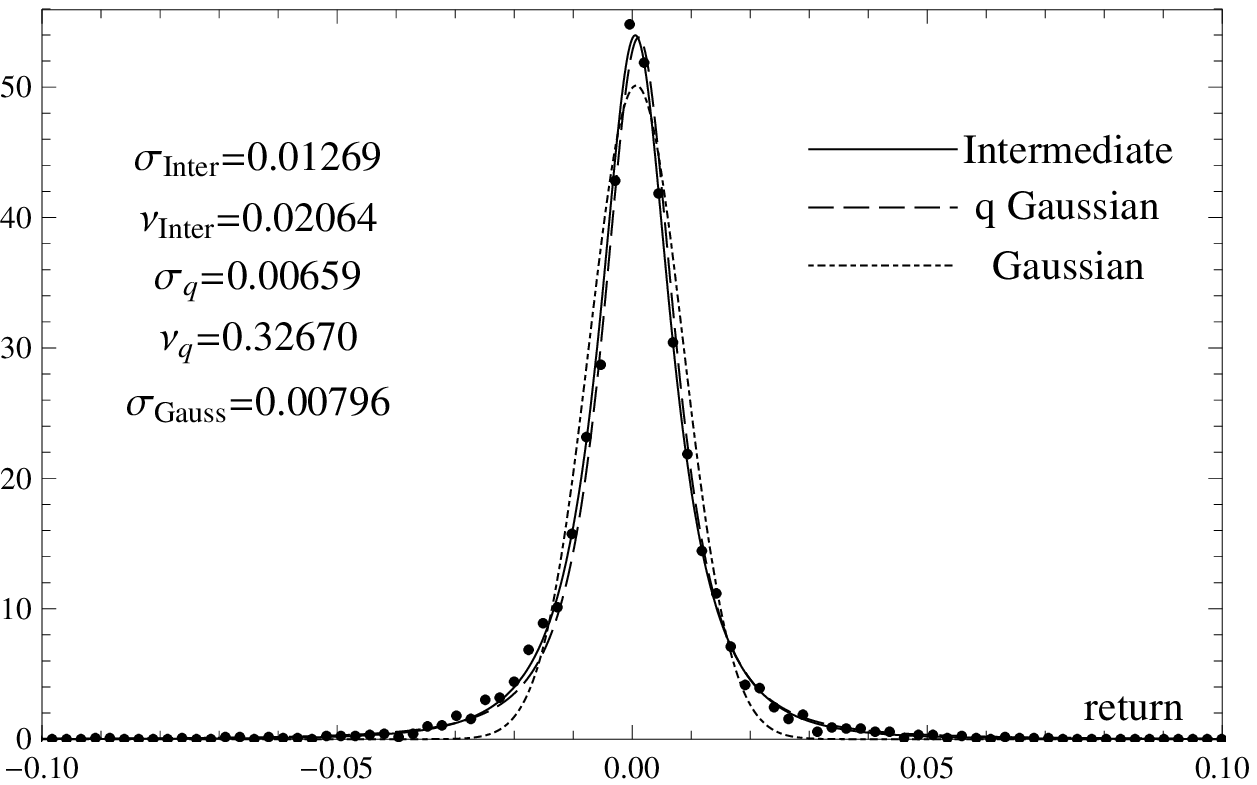} \hskip 1cm
\includegraphics[width=6.5cm]{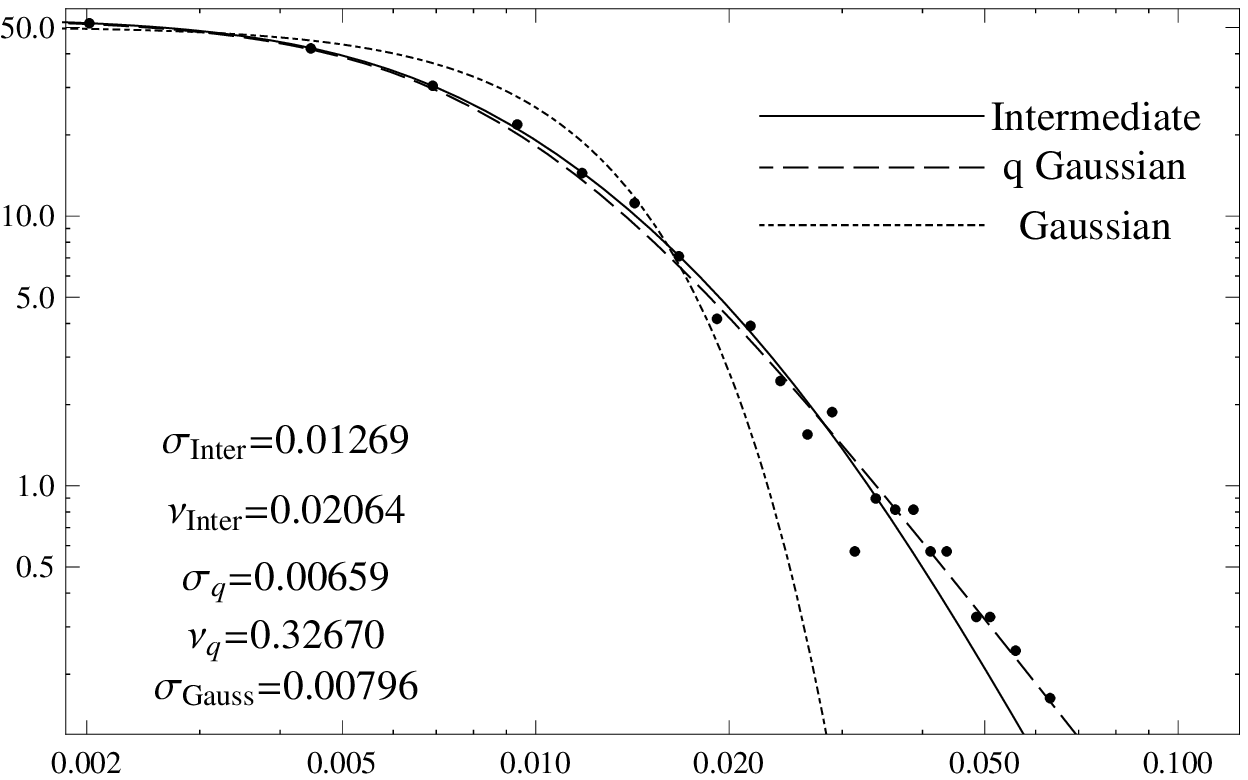}\newline
(a)~~~~~~~~~~~~~~~~~~~~~~~~~~~~~~~~~~~~~~~~~~~~~~~~~~~~~~~~~~~(b)
\end{center}
\caption{(a) The fitting of the market return of S\&P 500 price
index with the
intermediate, the q-Gaussian and the Gaussian distribution. (b) The corresponding log-log plot.}%
\end{figure}

\begin{figure}[ptb]
\begin{center}
\includegraphics[width=6.5cm]{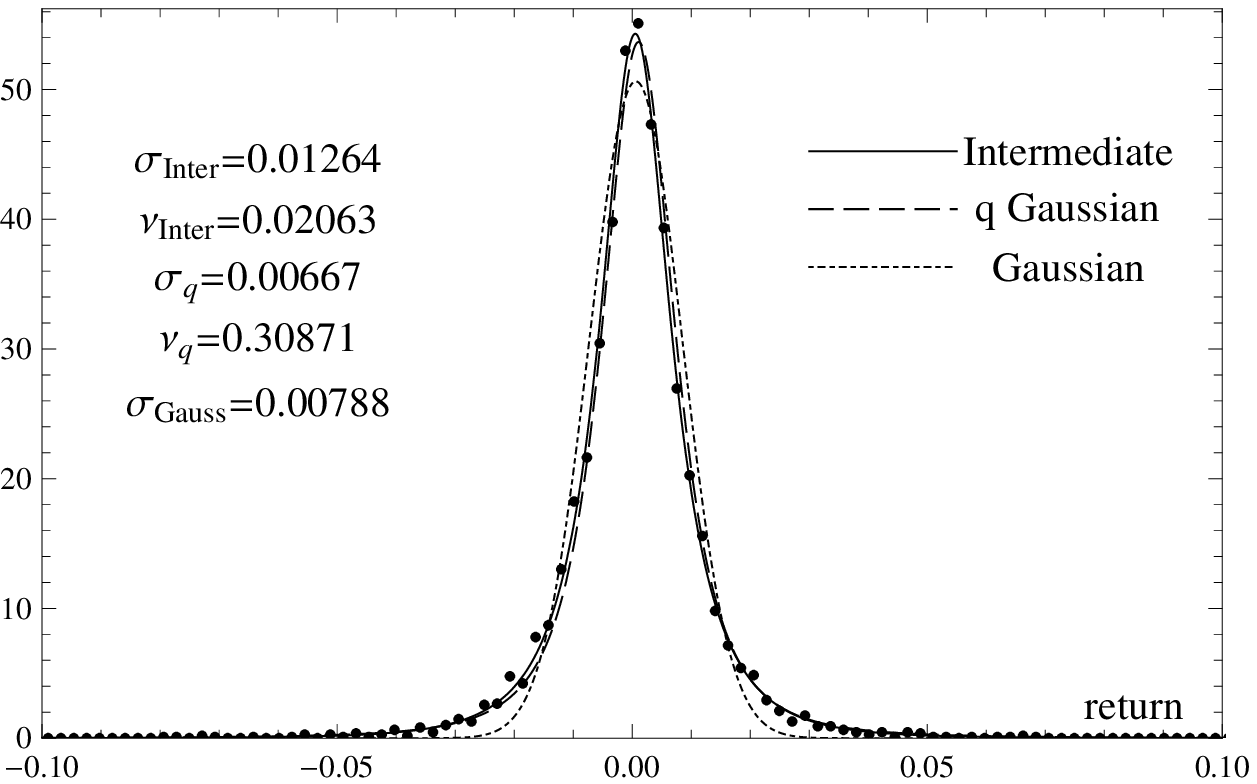} \hskip 1cm
\includegraphics[width=6.5cm]{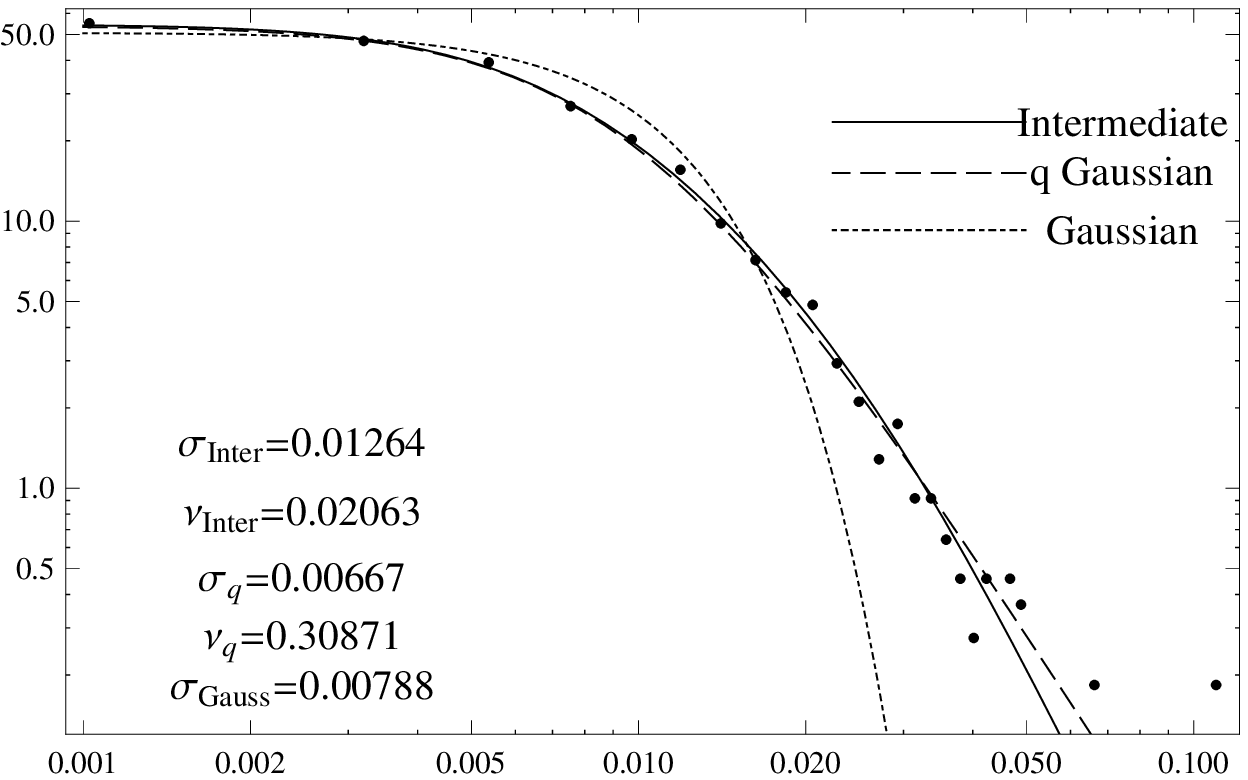}\newline
(a)~~~~~~~~~~~~~~~~~~~~~~~~~~~~~~~~~~~~~~~~~~~~~~~~~~~~~~~~~~~(b)
\end{center}
\caption{(a) The fitting of the market return of Dow Jones index
with the
intermediate, the q-Gaussian and the Gaussian distribution. (b) The corresponding log-log plot.}%
\end{figure}

\begin{figure}[ptb]
\begin{center}
\includegraphics[width=6.5cm]{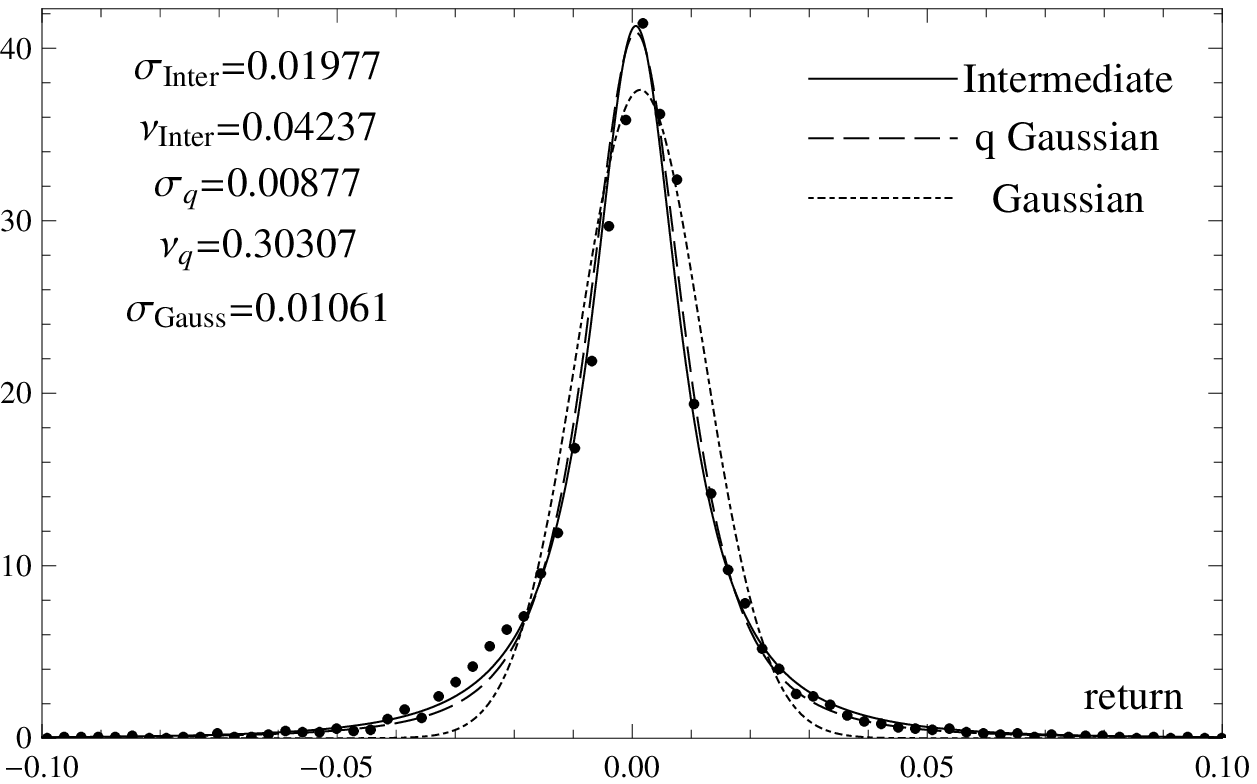} \hskip 1cm
\includegraphics[width=6.5cm]{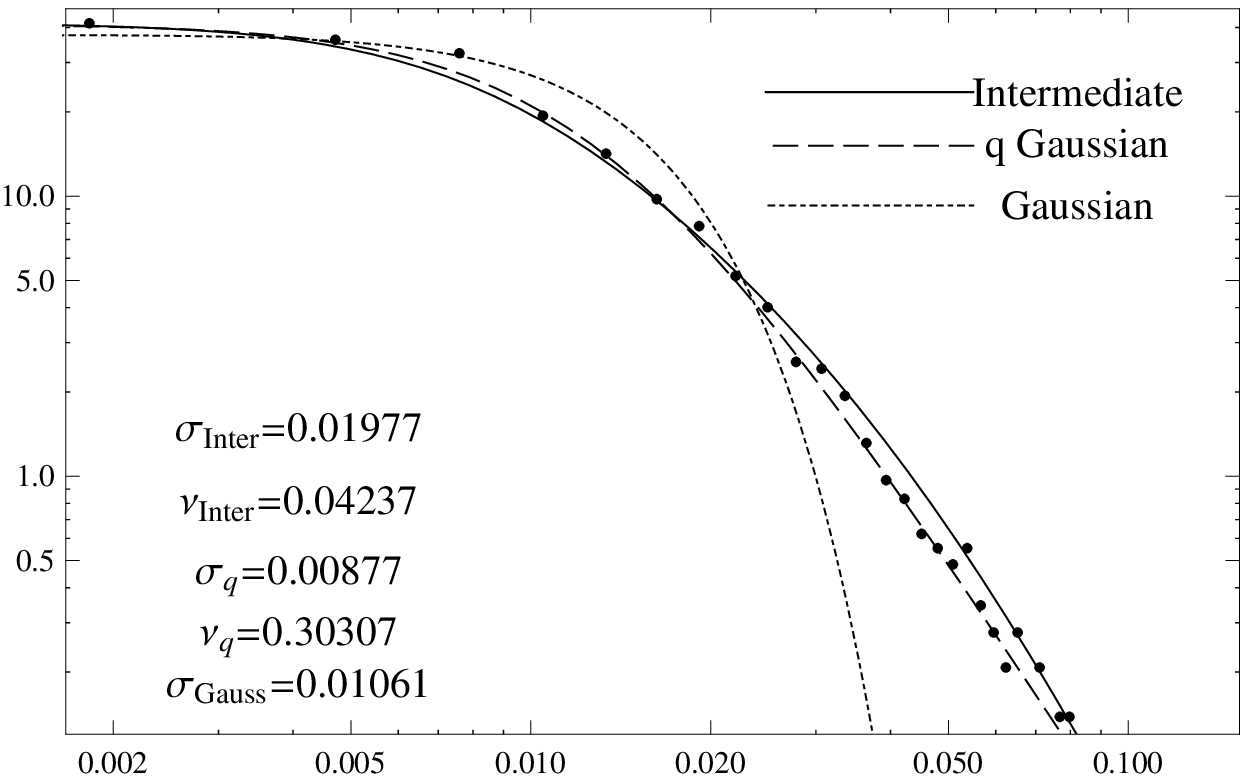}\newline
(a)~~~~~~~~~~~~~~~~~~~~~~~~~~~~~~~~~~~~~~~~~~~~~~~~~~~~~~~~~~~(b)
\end{center}
\caption{(a) The fitting of the market return of Nasdaq composite
index with the
intermediate, the q-Gaussian and the Gaussian distribution. (b) The corresponding log-log plot.}%
\end{figure}

\begin{figure}[ptb]
\begin{center}
\includegraphics[width=6.5cm]{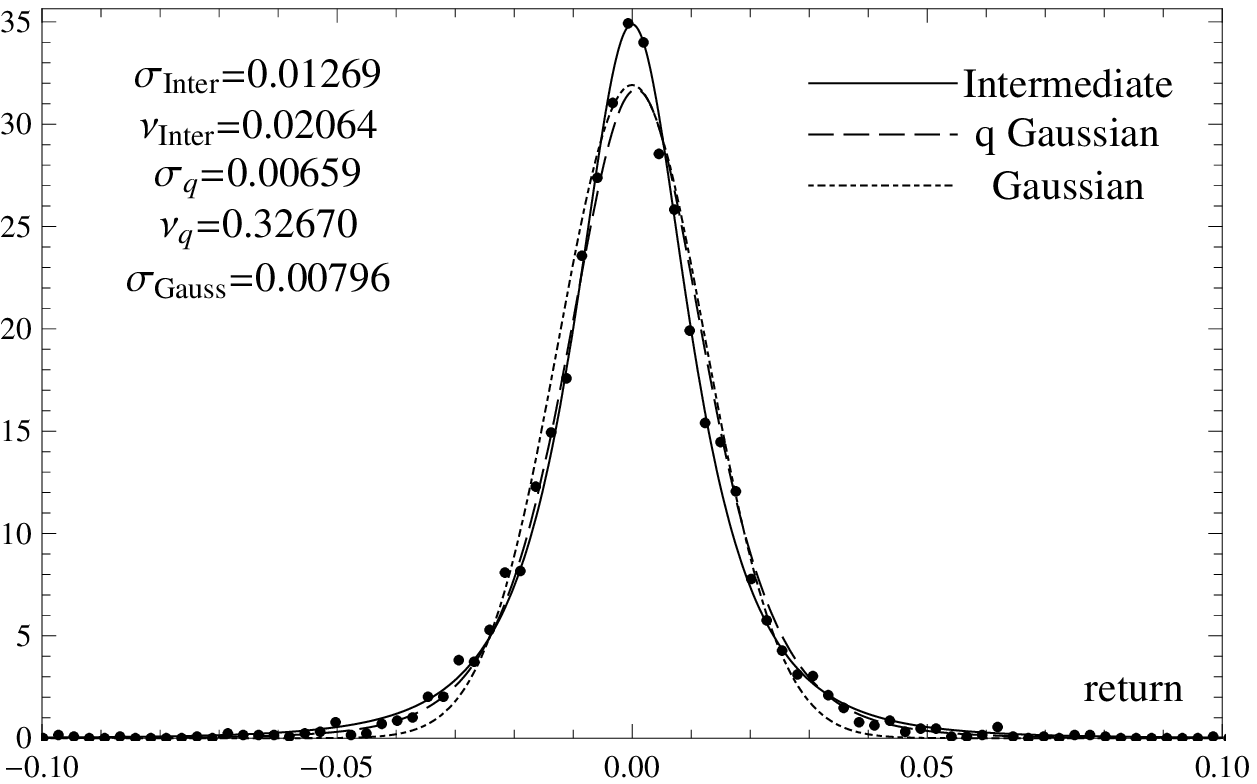} \hskip 1cm
\includegraphics[width=6.5cm]{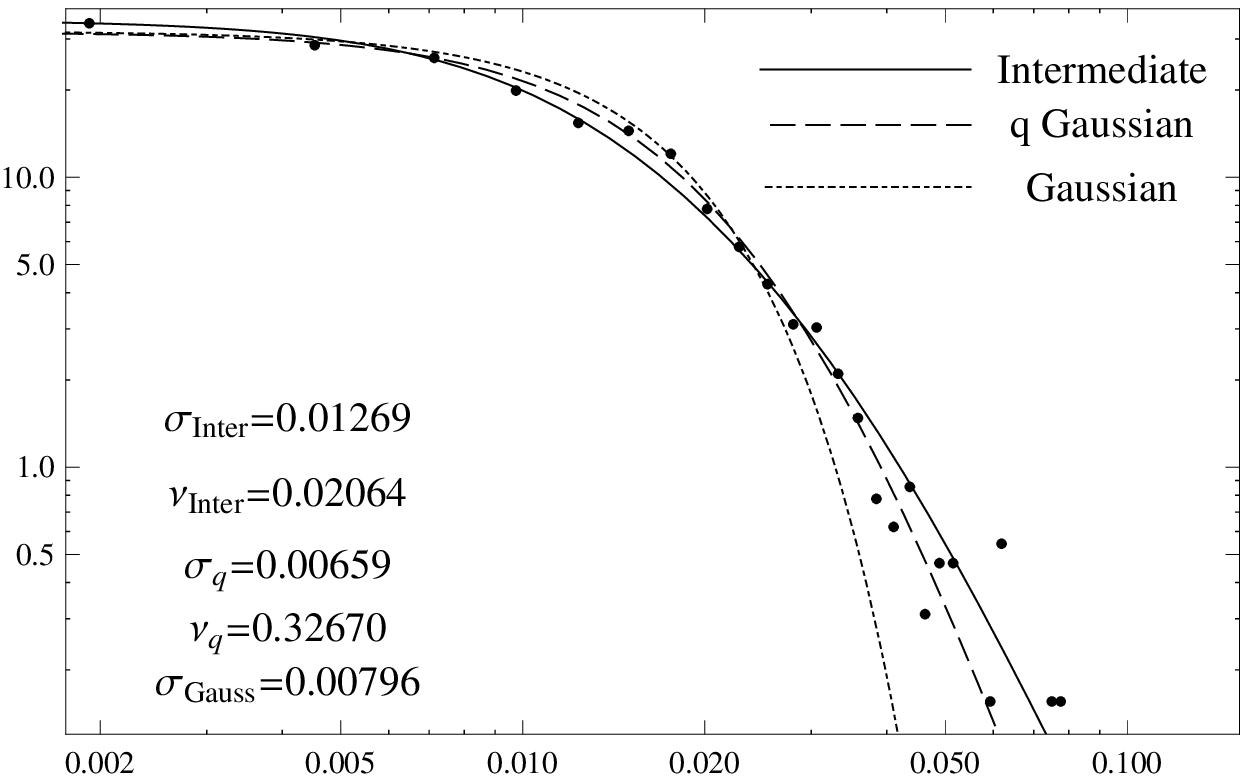}\newline
(a)~~~~~~~~~~~~~~~~~~~~~~~~~~~~~~~~~~~~~~~~~~~~~~~~~~~~~~~~~~~(b)
\end{center}
\caption{(a) The fitting of the market return of Nikkei 225 index
with the
intermediate, the q-Gaussian and the Gaussian distribution. (b) The corresponding log-log plot.}%
\end{figure}

\section{Conclusions and outlook}

In this paper, we construct an intermediate distribution between the Gaussian
and the Cauchy distribution. We provide the PDF and the corresponding
characteristic function of the intermediate distribution. The moment, on the
one hand, is used to show the mean value and the deviation to the mean value;
on the other hand, is used as the terms of a series of a quantity. Because
many kinds of distributions have no moment, we introduce the weighted moment.
We calculate the weighted moment with the cut-off and the exponential weighted
function, specifically. With the help of a smooth function, we give another
expression of the intermediate distribution.

As an application, the spectral line broadening in laser theory can be
approximated by a PDF. Since the homogeneous and inhomogeneous broadenings
exist simultaneously, we suggest to describe the line shape function by the
intermediate PDF, instead of the Cauchy or the Gaussian PDF. The intermediate
distribution is not only useful in spectral line broadening of laser but maybe
more attractive in financial physics \cite{adaptive estimation,moment
estimation,Maximum entropy estimation}. A number of theoretical and empirical
studies indicate that the majority of financial data are estimated as sharp
peak and fat tails PDFs, but the Gaussian distribution is not general enough
to account the excess kurtosis \cite{estimation function,estimates
variance,Maximum entropy autoregressive}. Our result shows that the
intermediate and q-Gaussian distribution, however, comparing with the Gaussian
distribution, are more suitable for fitting the financial data. The fitting
results show that the intermediate distribution is somewhat more suitable to
describe the kurtosis and the q-Gaussian distribution is more suitable to
describe the tails, respectively. Such results show that the intermediate
distribution, like the q-Gaussian distribution, can also be utilized in
financial physics. In further study, we can use the intermediate distribution
to price the option, and get the true value of the option. Option is one kind
of derivatives which is the core of pricing theory. We also can price other
derivatives whose price is based on the stock market return.

\vskip0.5cm \noindent\textbf{Acknowledgements} This work is supported in part
by NSF of China under Grant No. 11075115.


\begin{thebibliography}{99}                                                                                               %


\bibitem {GK}B.V. Gnedenko, A.N. Kolmogorov, 1949 Limit Theorems for Sums of
independent Random Variables, Translated from Russian by Addison-Wesley, 1968.

\bibitem {Samuels}S.M. Samuels, Positive-Integer-Valued Infinitely Divisible
Distribution, Department of Statistics, Division of mathematical Sciences,
Purdue University, 1975.

\bibitem {Novikov}E.A. Novikov, Phys. Rev. E 50 (1994) R3303.

\bibitem {Infinitely Divisible}K. Sato, Levy Processes and Infinitely
Divisible Distributions, Cambridge University Press, Cambridge, 1999.

\bibitem {Preda}V.C. Preda, Ann. Inst. Statist. Math. 34 (1982) 335.

\bibitem {reichl}L.E. Reichl, A Modern Course in Statistical Physics, 2nd edn,
Wiley, New York, 1998.

\bibitem {SW}Z.-S. She, E.C. Waymire, Phys. Rev. Lett. 74 (1995) 262.

\bibitem {Probability Analytic View}D.W. Stroock, Probability Theory: An
Analytic View, 2nd edn, Cambridge University Press, Cambridge, 2010.

\bibitem {TMP}C. Tsallis, R.S. Mendes, A.R. Plastino, Physica A 261 (1998) 534.

\bibitem {TRPB}C. Tsallis, A. Rapisarda, A. Pluchino, E.P. Borges, Physica A
381 (2007) 143 .

\bibitem {RST}A. Rodr\'{\i}guez, V. Schw\"{a}mmle, C. Tsallis, J. of Stat.
Mech. (2008) P09006.

\bibitem {DBR}P. Douglas, S. Bergamini, F. Renzoni. Phys. Rev. Lett. 96 (2006) 110601.

\bibitem {Bor}L. Borland, Phys. Rev. Lett. 89 (2002) 098701.

\bibitem {SB1}E. Van der Straeten, C. Beck, Phys. Rev. E 78 (2008) 051101.

\bibitem {SB2}E. Van der Straeten, C. Beck, Physica A, 390 (2011) 951.

\bibitem {Principles of Lasers}O. Svelto, Principles of Lasers, 5th edn,
Springer, New York, 2010.

\bibitem {Spinwave}W.-S. Dai, M. Xie, J. Stat. Mech. P04021 (2009).

\bibitem {ISA}Y. Shen, W.-S. Dai, M. Xie, Phys. Rev. A 75 (2007) 042111.

\bibitem {Gentile}W.-S. Dai, M. Xie, Ann. Phys. (NY) 309 (2004) 295.

\bibitem {GentileS}G. Gentile, Nuovo Cim. 17 (1940) 493.

\bibitem {GS}W.-S. Dai, M. Xie, J. Stat. Mech. P07034 (2009).

\bibitem {SU2}W.-S. Dai, M. Xie, Physica A 331 (2004) 497.

\bibitem {Pricing of Options}F. Black, M. Scholes, Journal of Political
economy 81 (1973) 637.

\bibitem {market equilibrium}W. Sharpe, Journal of Finance 19 (1964) 425.

\bibitem {valuation}J. Lintner, Review of Economics and Statistics 47 (1965) 13.

\bibitem {Equilibrium}J. Mossin, Econometrica 34 (1966) 768.

\bibitem {New}C. Tsallis, J. Stat. Phys. 52 (1988) 479.

\bibitem {Levy Processes}D. Applebaum, Levy Processes and Stochastic Calculus,
2nd edn, Cambridge University Press, Cambridge, 2009.

\bibitem {S. S. Chern}S.S. Chern, W.H. Chen, K.S. Lam, Lectures on
differential geometry,World Scientific, Singapore, 1999.

\bibitem {S. P. Novikov}B.A. Dubrovin, A.T. Fomenko and S.P. Novikov, Modern
Geometry --- Methods and Applications, Part 2, Springer, Berlin, 1985.

\bibitem {Effect Laser Fano}U. Fano, Phys. Rev. 124 (1961) 1866.

\bibitem {Laser Experiment}M. Hanif, M. Aslam, M. Riaz, S.A. Bhatti, M.A.
Baig, J. Phys. B 38 (2005) S65.

\bibitem {Handbook of Laser}C.E. Webb, J.D.C. Jones, Handbook of Laser
Technology and Applications, Volume I, IOP Publishing, London, 2004.

\bibitem {spectral data}J.Rosato, D. Boland, M.Difallah, Y.Marandet, R. Stamm,
International Journal of Spectroscopy (2010) Article ID 374372.

\bibitem {New1}A.-H. Sato, J. Phys.: Conf. Ser. 201 (2010) 012008.

\bibitem {New2}A.-H. Sato, Phys. Rev. E 69 (2004) 047101.

\bibitem {New3}V. Gontis, J. Ruseckas, A. Kononovi\v{c}ius, Physica A 389
(2010) 100.

\bibitem {New5}M. Gell-Mann, C. Tsallis, Nonextensive Entropy
Interdisciplinary Applications, Oxford University Press, New York, 2004.

\bibitem {New6}H. Suyari, Physica A 368 (2006) 63.

\bibitem {adaptive estimation}J.B.\ McDonald, W.K. Newey, Econometrics Theory
4 (1988) 428.

\bibitem {moment estimation}R.J. Smith, The Economic Journal 107 (1997) 503.

\bibitem {Maximum entropy estimation}H.K. Ryu, Journal of Econometrics 56
(1993) 397.

\bibitem {estimation function}D.X. Li, H.J. Turtle, Journal of Business and
Economic Statistics 18 (2000) 174.

\bibitem {estimates variance}R.F. Engle, Econometrica 50 (1982) 987.

\bibitem {Maximum entropy autoregressive}S.Y. Park, A.K. Bera, Journal of
Econometrics 150 (2009) 219.
\end{thebibliography}
\end{document}